\documentclass[
 	reprint,
	amsmath,
    amssymb,
	aps,
    floatfix,
    superscriptaddress
]{revtex4-1}

\usepackage[english]{babel}
\usepackage[utf8x]{inputenc}
\usepackage[T1]{fontenc}
\usepackage{braket}

\usepackage{amsmath}
\usepackage{amsfonts}
\usepackage{graphicx}
\usepackage{dcolumn}
\usepackage{bm}
\usepackage[colorinlistoftodos]{todonotes}
\usepackage[colorlinks=true, allcolors=blue]{hyperref}

\usepackage{soul}
\setstcolor{red}

\begin{document}

\title{Electronic properties of correlated kagom\'e metals AV$_3$Sb$_5$ (A = K, Rb, Cs): A perspective}
\author{Thanh Nguyen}
\affiliation{Department of Nuclear Science and Engineering, Massachusetts Institute of Technology, Cambridge, MA 02139, USA}

\author{Mingda Li}
\thanks{Corresponding author.\\\href{mailto:mingda@mit.edu}{mingda@mit.edu} \vspace{0.5cm}}
\affiliation{Department of Nuclear Science and Engineering, Massachusetts Institute of Technology, Cambridge, MA 02139, USA}

\date{\today}

\begin{abstract}
    Following the discovery of a new family of kagom\'e prototypical materials with structure AV$_3$Sb$_5$ (A = K, Rb, Cs), there has been heightened interest in studying the correlation-driven electronic phenomena in these kagom\'e lattice systems. The study of these materials has gone beyond magneto-transport measurements to reveal exciting features such as Dirac bands, anomalous Hall effect, bulk superconductivity with $T_c$ $\sim$ 0.9 K-2.5 K, and the observation of charge density wave instabilities, suggesting an intertwining of topological physics and new quantum orders. Moreover, very recent works on numerous types of experiments have appeared further examining the unconventional superconductivity and the exotic electronic states found within these kagom\'e materials. Theories on the strong interactions that play a role in these systems have been proposed to shed light on the nature of these topological charge density waves. In this brief review, we summarize these recent experimental findings and theoretical proposals, and envision the materials as new platforms to study the interplay between topological physics and strongly-correlated electronic systems.
\end{abstract}

\maketitle


\section{Introduction} \label{sec:introduction}
Materials with kagom\'e crystal structures have attracted significant interest due to the emergence of flat bands with electronic band structures that can host Dirac cones and van Hove singularities. Various exotic phenomena can emerge from the interplay between nontrivial band topology and strong Coulomb correlations. Formation of superconducting ground states in layered kagom\'e compounds is rare, therefore the discovery by groups at the University of California, Santa Barbara and at the Colorado School of Mines, of a new family of kagom\'e metals, AV$_3$Sb$_5$ (A = K, Cs, Rb) has led to a plethora of substantial findings \cite{ortiz2019}. Among these include the manifestation of superconductivity at 0.9 K-2.5 K and an intriguing charge-density-wave transition around 78 K-103 K. These kagom\'e metals have uncovered a fascinating platform for new insights into the rich physics that lies at the interplay between strong electronic correlations, nontrivial band topology, unconventional superconductivity, and emergent quantum orders. 

This brief review is organized as follows. In Section \ref{sec:synthesis}, we describe the crystal structure and synthesis procedure of these kagom\'e systems. Section \ref{sec:cdw-superconductivity} is a discussion of the experimental results in connection with the competition between the charge density wave (CDW) ordering and the presumably unconventional superconductivity by means of measurements performed at ambient and high pressures. Section \ref{sec:bandtopology} provides a synopsis of the experimental characterizations of the nontrivial band topology of the material system. We expand in Section \ref{sec:anomalous} on measurements of the large anomalous transport in these systems in addition to experimental findings of several coexisting orders from scanning tunneling microscopy (STM) (Section \ref{sec:STM}). We discuss some of the ongoing efforts on the theoretical side to explain the phenomena observed in these systems in Section \ref{sec:theory}. We conclude with a perspective on the implications that derive from these experimental and theoretical studies on our understanding of strongly-correlated topological kagom\'e superconductors as well as outstanding questions and potential applications using this family of materials.

\section{Crystal Structure and Synthesis} \label{sec:synthesis}
AV$_3$Sb$_5$ crystallizes in the hexagonal P6/$mmm$ (191) space group (Fig. \ref{fig:1}a) where the vanadium (V) sublattice forms a two-dimensional kagom\'e net that is interwoven with a hexagonal net of antimony (Sb) atoms. This V-Sb layer is sandwiched between two honeycomb layers of Sb atoms which form a different sublattice from the interwoven Sb atoms and altogether, these are intercalated by a hexagonal net of alkali (A) atoms. Due to the layered nature of the crystal structure, the crystals lend themselves to be exfoliable to thin 2D layered sheets.\\

The original synthesis technique \cite{ortiz2019, ortiz2020} consists of ball-milling the different constituent elements into grounded powders, sieving the powders through meshes and synthesizing the crystal via the self-flux method. The best condition for the molar ratio between the constituent elements varies between different groups. The flux consists of a eutectic mixture of VSb$_2$ with KSb$_2$ and KSb (for KV$_3$Sb$_5$); with RbSb and RbSb$_2$ or Rb$_3$Sb$_7$ (for RV$_3$Sb$_5$); or with CsSb and CsSb$_2$ or Cs$_3$Sb$_7$ (for CsV$_3$Sb$_5$). The powders are sealed into alumina crucibles under argon gas and stainless steel jackets. After quickly heating to a $\sim$1000$^\circ$C followed by a 24-hour soaking period and one or two slow cooling stages at lower temperatures (to 800-900$^\circ$ at 100$^\circ$C/hr and/or to 200-650$^\circ$C at 1.5-3$^\circ$C/hr), the excess flux is removed by water etching. Special care must be taken due to the high reactivity of the alkali metals; therefore, the powders are purified to remove residual oxides (particularly for vanadium) and mixed in an argon glove box with a slight extra amount of the alkali to compensate for its reactivity. The resulting millimetre-sized AV$_3$Sb$_5$ single crystals resemble shiny hexagonal flakes with a metallic luster and are stable in air.

Subsequent reports \cite{ortiz2021} reveal the necessary combination of the steel jackets, of the mechanical extraction and of the powder purification to obtain single crystals deficient of the alkali metal. However, many studies have no mention of the steel jackets and opt to instead decant the ampoule using a centrifuge to separate out the flux; the properties of the obtained crystals remain consistent with existing literature \cite{yin2021}. Surface treatment of the raw materials in a hydrogen atmosphere can also be performed. Some groups use magnesium oxide or tantalum crucibles instead of alumina ones with similarly obtained results. Thin flakes with nanometer-thicknesses can be exfoliated from the single crystals using conventional Scotch tape or using an Al$_2$O$_3$-assisted exfoliation method. These flakes are prone to oxidation once exposed to air.

\section{Interplay between CDW Order and Superconductivity} \label{sec:cdw-superconductivity}
\begin{figure}[ht!]
	\centering
	\includegraphics[width=\linewidth]{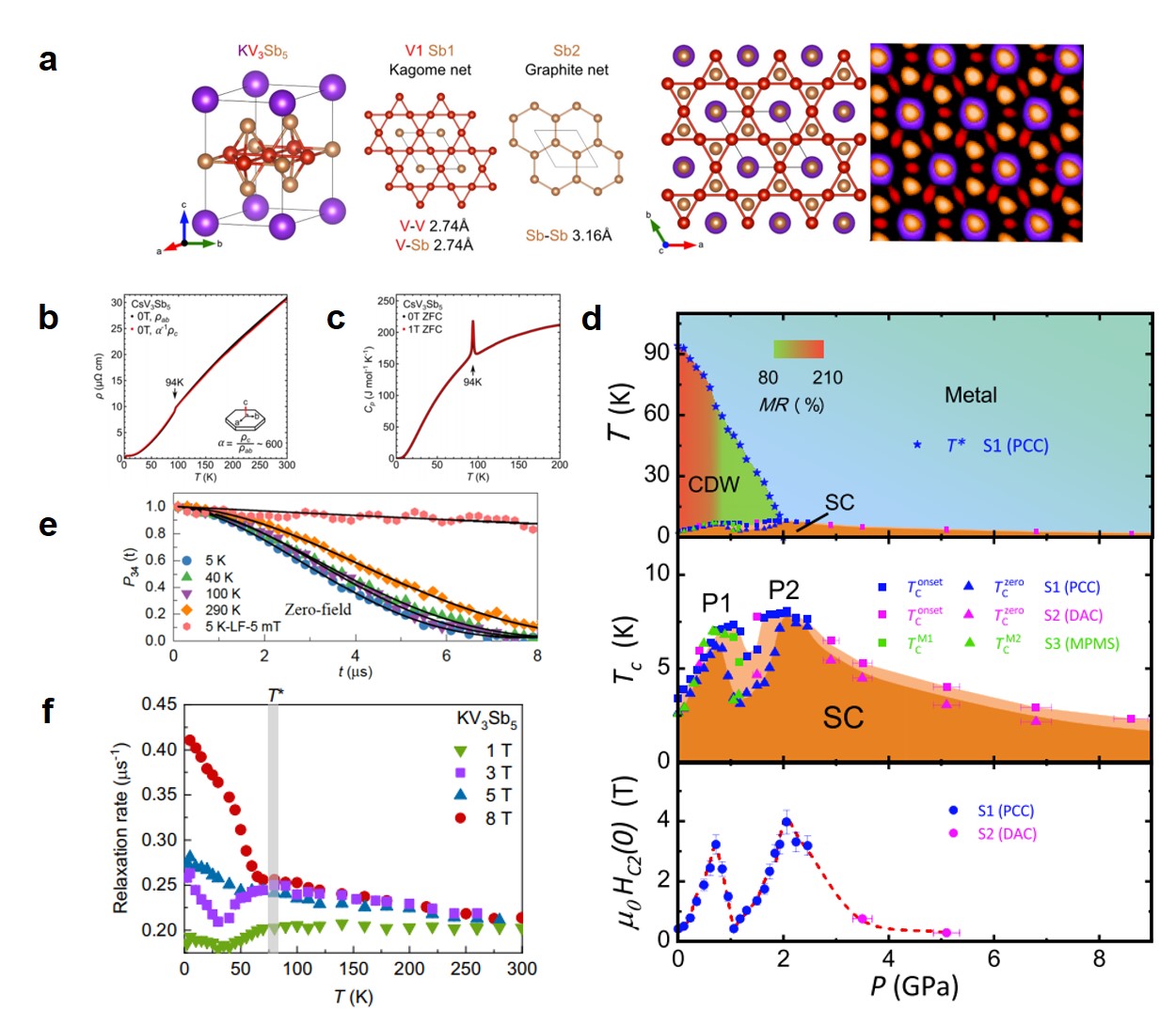}
	\caption{\textbf{Crystal structure, double-dome superconductivity, and $\mu$SR measurements.} \textbf{a)} Crystal structure of AV$_3$Sb$_5$ formed by a vanadium (V) kagom\'e net, two distinct antimony (Sb) sublattices, intercalated by a layer of alkali atoms. \textbf{b)} Electrical resistivity and \textbf{c)} heat capacity measurements of CsV$_3$Sb$_5$ demonstrating an anomaly near 94 K associated with charge ordering. \textbf{d)} Pressure-temperature phase diagram of CsV$_3$Sb$_5$ displaying charge density wave ordering (CDW) and superconductivity (SC) (top). Plots of the superconducting transition temperature $T_c$ (middle) and of the upper critical field at zero temperature $\mu_0 H_{C2}(0)$ (bottom) as a function of pressure both display features of double-dome superconductivity. PCC: piston cylinder cell; DAC: diamond anvil cell; MPMS: MPMS-5T, Quantum Design. \textbf{e)} Zero field $\mu$SR time spectra of KV$_3$Sb$_5$ at different temperatures with black lines serving as fits to a depolarization function. \textbf{f)} Plot of the muon spin relaxation rate as a function of temperature under the $c$-axis at different magnetic fields. The grey line represents the charge ordering temperature $T^{*}$. \textbf{a-c} from Ref. \cite{ortiz2020}, \textbf{d)} from Ref. \cite{yu2021(2)}, and \textbf{e-f} from Ref. \cite{mielke2021}.}
	\label{fig:1}
\end{figure}

Unlike a few other kagom\'e systems such as Co$_3$Sn$_2$S$_2$ \cite{liu2018,wang2018,yin2019,liu2019,morali2019}, Mn$_3$Sn \cite{nakatsuji2015,kuroda2017}, Mn$_3$Ge \cite{nayak2016}, Fe$_3$Sn$_2$ \cite{ye2018,yin2018,lin2018}, FeSn \cite{kang2020}, and TbMn$_6$Sn$_6$ \cite{yin2020}, the AV$_3$Sb$_5$ family does not exhibit detectable magnetic ordering down to the lowest measurable temperatures from neutron diffraction measurements.This behaviour is reminiscent of other kagom\'e compounds such as CoSn \cite{liu2020,kang2020(2)}, quantum spin liquid candidates herbertsmithite \cite{han2012,fu2015,norman2016} and Zn-substituted barlowite \cite{feng2017}, and metal-organic frameworks \cite{barreteau2017,jiang2021(2)}. On the other hand, bulk electric transport, heat capacity, and magnetization measurements (Figs. \ref{fig:1}b, c for CsV$_3$Sb$_5$) \cite{ortiz2019, ortiz2020} additionally reveal the presence of an anomaly near 78 K-103 K for all three members of the family ($T_{\text{CDW}}$ = 78 K (K), 94 K (Cs), 103 K (Rb)), due to the formation of a charge order with the presence of a charge density wave (CDW) instability. Further details on STM measurements of the CDWs and related instabilities are discussed in Section \ref{sec:STM}.

The observation of bulk superconductivity in single crystals of AV$_3$Sb$_5$ has been reported with a superconducting transition temperature of $T_c$ = 0.93 K (K), 2.5 K (Cs), 0.92 K (Rb) \cite{ortiz2020, ortiz2021, yin2021} with measurements suggesting that the superconductivity is in the strong-coupling regime ($2\Delta/k_BT_C = 5.2$). In kagom\'e lattices, unconventional superconductivity can emerge by instabilities driven by interactions owing to a high level of nesting of the Fermi surface whereupon scattering occurs between saddle points of the band at the M (TRIM) points in the Brillouin zone. Indeed, in Josephson junctions formed with K$_{1-x}$V$_3$Sb$_5$ \cite{wang2020}, measurements of the magnetoresistance reveal spin-polarized supercurrents and those of the current versus phase contain anomalous interference patterns with a minimum at zero field (when sweeping an out-of-plane field), both of which would provide possible evidence of spin-triplet superconductivity. These bits of evidence for unconventional superconductivity are accompanied by an observation of edge-localized supercurrents which are in line with DFT calculations of topological edge states. The nontrivial topological band structures may also have some ramifications for the stabilization of Majorana zero modes within the vortex core of proximitized superconducting surface states, as the case for the superconducting Dirac surface states in Bi$_2$Te$_3$/NbSe$_2$ heterostructures. There is evidence of a robust zero-bias conductance peak measured using STM inside the superconducting vortex core on the Cs surface which would resemble a Majorana bound state \cite{liang2021}.

A stronger form of evidence for unconventional superconductivity in these compounds originates from high-pressure studies up to 100 GPa \cite{zhao2021, yu2021(2), chen2021, du2021, zhang2021, chen2021(3), zhu2021, wang2021(5), du2022}. As shown in Fig. \ref{fig:1}d for CsV$_3$Sb$_5$ (but also the case for the other members), high-pressure measurements of the thermal conductivity, magnetic susceptibility and electrical resistivity reveal two superconducting domes. The first dome is suggested to be related to the suppression of CDW-instability mentioned previously, whereas the second dome may present another exotic pairing mechanism. The presence of an interval of pressure which exhibits no superconductivity and the re-emergence of superconductivity as a second phase (typically called SC-II) would suggest nodal superconductivity, and it is very reminiscent of the behavior of the high-temperature cuprate superconductors. There is also an intimate competition between the CDW-like order and the superconductivity as revealed by the fact that the CDW transition temperature ($T_{\text{CDW}}$) decreases with pressure before vanishing at $\sim$ 2 GPa in a first-order like manner while the second superconducting dome persists up to $\sim$ 100 GPa. The re-emergence of the second dome is not well-understood as the crystal structural stability of CsV$_3$Sb$_5$ seems to persist at high pressures  with some suggesting that it may be attributed to a pressure-induced Lifshitz transition resulting from strengthened electron-phonon coupling due to a change in the electronic structure. Ensuing synchrotron x-ray diffraction measurements of the second dome in KV$_3$Sb$_5$ and RbV$_3$Sb$_5$ (but not of CsV$_3$Sb$_5$) \cite{du2021(2)} show that the minimum of $T_c$ between the domes of these materials is associated with structural phase transition from hexagonal to monoclinic structures. In KV$_3$Sb$_5$ particularly, there is a subsequent transition to an orthorhombic structure which coincides with the peak of the second dome. Another high-pressure x-ray diffraction measurement showcases a crossover from two-dimensional to three-dimensional behavior as the interlayer Sb-Sb bonds form which leads to the increase in $T_c$ for the second dome \cite{yu2021(4)}. Furthermore, the superconductivity demonstrates anisotropy of the upper and lower critical fields ($H_{c2}$ and $H_{c1}$, respectively) along the $a$-axis versus along the $c$-axis, indicating strong anisotropic scatterings that weave in with the coexisting rotation-symmetry breaking orders present in the system \cite{ni2021, xiang2021}. These features are also revealed upon application of uniaxial strain \cite{qian2021}. Soft and mechanical point-contact spectroscopy measurements show how local strain induced by point contacts enhances the superconductivity in KV$_3$Sb$_5$ and CsV$_3$Sb$_5$ \cite{yin2021(2)}. There is no full consensus on the unconventional superconductivity by far. For instance, one study (among many in this review) of the magnetic penetration depth using a tunneling diode oscillator \cite{duan2021} suggests that the superconductivity is described by a two-gap s-wave nodeless model whereas other studies might imply single-gap and/or nodal superconductivity.

A multitude of $\mu$SR measurements delve deeper into the nature of the superconductivity. Refs. \cite{gupta2021, mielke2021} reports the presence of two s-wave gaps at the Fermi surface with an anisotropy of the temperature dependence of the magnetic penetration depth $\lambda$ with a large value of $T_c/\lambda^{-2}(0)$ suggesting a unconventional pairing mechanism akin to hole-doped cuprates. However, unlike Ref. \cite{gupta2021}, the zero-field measurements of Refs. \cite{mielke2021, yu2021(3)} report a clear signature of a time-reversal symmetry breaking charge order emerging from the in-plane 2$\times$2 CDW phase which may be explained via a theorized chiral flux phase (Figs. \ref{fig:1}e-f). The emergence of this signal at $\sim$70 K does not exactly coincide with the CDW transition temperature ($\sim$94 K) in the $\mu$SR studies. Time-resolved optical Kerr spectroscopy measurements, a powerful optical method to probe time-reversal symmetry breaking state, indicate that this feature along with the two-fold rotation symmetry do indeed develop simultaneously at the transition temperature \cite{wu2021(2)}.

Carrier doping of the AV$_3$Sb$_5$ kagom\'e metals would allow the tunability of the van Hove singularities (vHs) at the M points and the electron pocket at the $\Gamma$ point relative to the Fermi level. Studies of carrier doping on these compounds would elucidate more on the mechanisms behind the CDW, the superconductivity and the coupling between these two, as well as the the stability of these phase transitions. A study on CsV$_3$Sb$_{5-x}$Sn$_x$ (with 0 $\leq$ x $\leq$ 1.5) suggests an intertwining between the superconducting and CDW orders \cite{oey2021} with Sn doping which introduces hole carriers. As the doping increases, the superconducting transition temperature evolves non-monotonically with a two-dome feature and at the point where the vHs moves through the Fermi level at the M-point, there is an accompanied suppression of CDW order. A study on Cs(V$_{1-x}$Ti$_x$)$_3$Sb$_5$ (0 $\leq$ x $\leq$ 0.117), which introduces hole carriers via Ti doping, show similar behavior of enhanced superconductivity with doping, competition between CDW and superconductivity, as well as the absence of CDW ordering once the vHs is moved above the Fermi level \cite{liu2021(2)}. A similar study of Ti-doping reveals two distinct superconducting phases that differ in the shape of gap pairing \cite{yang2021(3)}. Although the CDW order finds its origins to the vHs near the $M$ point and the superconductivity, to the Sb electron pocket at the $\Gamma$ point, there seems to be a competition between these two orders. ARPES studies of Cs-doped CsV$_3$Sb$_5$ show how electron doping can drastically alter the CDW properties leading to suppression of the CDW phase \cite{nakayama2022}. Measurements have also been performed on exfoliated thin flakes of CsV$_3$Sb$_5$ \cite{song2021(2), song2021(3), wang2021(4)} to investigate the superconductivity and CDW with varied sample thicknesses. They reveal that the superconducting transition temperature and the upper critical field of thin flakes are enhanced compared to bulk samples, which is accompanied by the suppression of the CDW ordering. The change with sample thickness is non-monotonic, which may be explained by a 3D-to-2D crossover at a critical thickness of around 60 nm. This further supports the idea of a competition between CDW and superconductivity in the system. In particular, Ref. \cite{song2021(2)} showcases how hole doping at the Cs site enhances $T_c$ and suppresses the CDW order.

A myriad of other experimental techniques have also been applied to further investigate the CDW state in this kagom\'e metal family. Hard x-ray scattering measurements demonstrate that the CDW may have a three-dimensional character (a 2$\times$2$\times$2 superstructure) \cite{li2021}. These measurements also find that the CDW is not driven by strong electron-phonon coupling as there is an absence of corresponding acoustic phonon anomalies, but rather the CDW may be of an unconventional nature as a result of kagom\'e lattice nesting at specific vHs filling, consistent with ARPES measurements described in Section \ref{sec:bandtopology}. On the other hand, neutron scattering experiments on CsV$_3$Sb$_5$ in Ref. \cite{xie2021} propose that electron-phonon coupling must play an important role for the CDW order. While the study does not observe a dramatic change in the signals for acoustic phonons (which is consistent with hard x-ray scattering measurements), there is an observation of a static lattice distortion and a hardening of a longitudinal optical phonon which may be associated with the $2\times2$ charge order and the inverse star of David CDW structure. In this case, the CDW would not have a purely electronic origin from instabilities related to the vHs, but instead electron-phonon coupling must play an important role. Pump-probe measurements reveal that the excited quasiparticle relaxation dynamics is explained by a CDW condensate that would follow a second order phase transition \cite{wang2021(3)}. Nuclear magnetic resonance (NMR) measurements from Ref. \cite{song2021} suggests a three-dimensional nature of the CDW phase and the presence of dominant orbital order. Those from Ref. \cite{mu2021, mu2022} which include nuclear quadrupole resonance (NQR) measurements show similar results while further reporting a first-order nature of the CDW transition and evidence of s-wave superconductivity in CsV$_3$Sb$_5$. Ultrafast coherent phonon spectroscopy measurements of the CDW order \cite{ratcliff2021} complement the STM measurements of uniaxial order at low temperatures. Polarized Raman spectroscopy of CsV$_3$Sb$_5$ uncovers a phonon anomaly below ($\sim$50 K) and at the CDW transition temperature ($\sim$94 K) which is attributed to secondary electronic instabilities in the CDW state with s-wave symmetry \cite{wulferding2021}. Thickness-dependent Raman spectroscopy measurements on CsV$_3$Sb$_5$ \cite{liu2022} demonstrate enhanced changes in phonon properties as the thickness decreases which highlights large electron-phonon coupling. More recently, studies from temperature-dependent x-ray diffraction \cite{stahl2021, xiao2022}, polarization-resolved electronic Raman spectroscopy \cite{wu2022}, and ARPES \cite{li2021(3)} reveal the existence of a 2$\times$2$\times$4 superstructure in CsV$_3$Sb$_5$ at higher temperatures in contrast to the 2$\times$2$\times$2 superstructure at lower temperatures. Although the temperature values do not agree between these studies, the evidence of this superstructure reveals a change in the stacking of the V$_3$Sb$_5$ layers along the $c$ axis which may explain the two successive phase transitions above $\sim$60K as well as different temperature scales in the magneto-transport data.

\section{Nontrivial Band Topology} \label{sec:bandtopology}
\begin{figure}[h!]
	\centering
	\includegraphics[width=\linewidth]{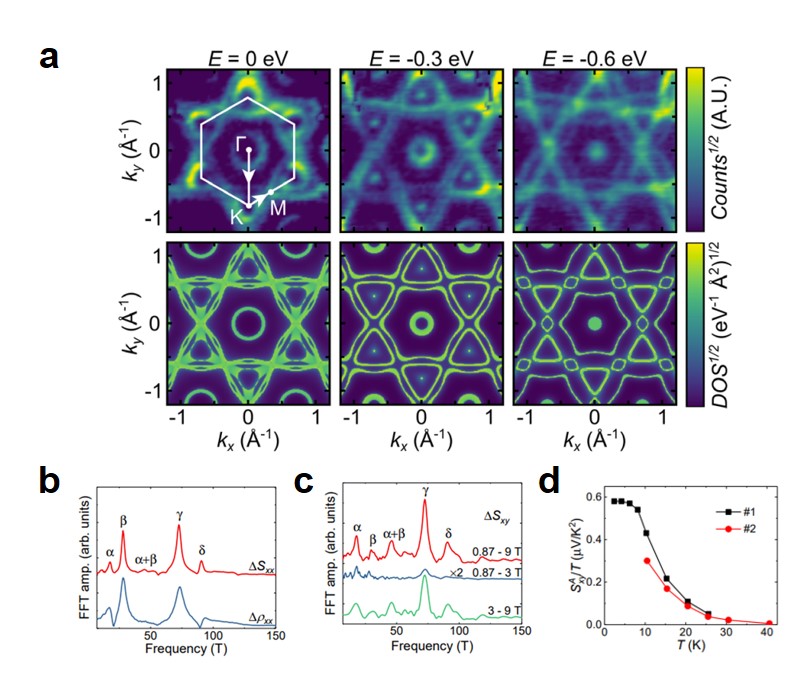}
	\caption{\textbf{ARPES, quantum oscillations, and anomalous Nernst effect.} \textbf{a)} ARPES measurements displaying the hexagonal Brillouin zone in addition to excellent agreement with density functional theory (DFT). Fast Fourier transform spectra of the oscillations from the longitudinal Seebeck signal (red in \textbf{b)}), from the longitudinal resistivity (blue in \textbf{b)}) and from the Nernst signal (in \textbf{c)}, at different ranges of magnetic field) of CsV$_3$Sb$_5$. \textbf{d)} Anomalous Nernst signals (divided by temperature) plotted as a function of temperature for two samples (called \#1 and \#2). \textbf{a} from Ref. \cite{ortiz2020}, \textbf{b-d} from Ref. \cite{chen2021(4)}}
	\label{fig:2}
\end{figure}

Angle-resolved photoemission spectroscopy (ARPES) measurements (Fig. \ref{fig:2}a) and quantum oscillations \cite{yin2021,fu2021,shrestha2022} reveal the presence of multiple Dirac points near the Fermi surface and with corresponding topological protected surface states that occur at the M points (a time-reversal invariant momentum (TRIM) point) in the Brillouin zone in addition to small Fermi surfaces with low effective mass and non-zero Berry phases. This is found to be consistent with calculations of the electronic band structure with density functional theory (DFT), where one also identifies the existence of these features. When one calculates the $\mathbb{Z}_2$ topological invariant between each pair of bands near the Fermi level (both time-reversal and inversion symmetries are present in the system), one categorizes the family of kagom\'e metals as $\mathbb{Z}_2$ topological metals thereby identifying them with nontrivial band topology. There have also been ARPES observations of Dirac nodal lines and loops near the Fermi level along the $k_z$ direction and surrounding the H points \cite{hao2021}.
\begin{figure*}[ht!]
	\centering
	\includegraphics[width=\linewidth]{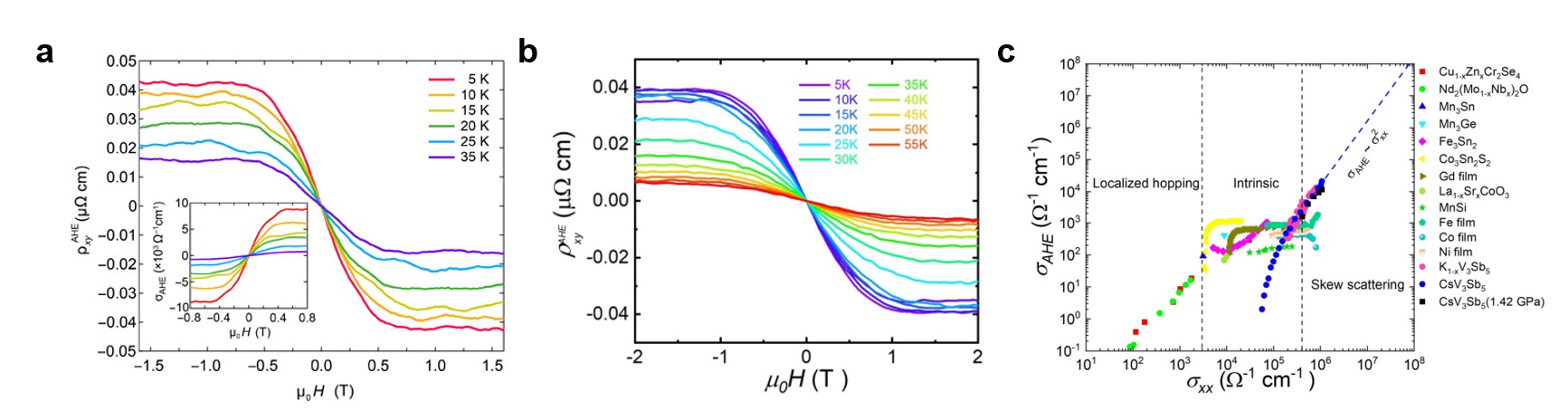}
	\caption{\textbf{Large AHE.} Extracted values of $\rho_{xy}^{\text{AHE}}$ for \textbf{a)} KV$_3$Sb$_5$ (with $\sigma_{xy}^{\text{AHE}}$ as an inset) and \textbf{b)}  CsV$_3$Sb$_5$ as a function of magnetic field. \textbf{c}) The authors of these AHE results argue that the behavior of $\sigma_{\text{AHE}}$ versus $\sigma_{xx}$ of both kagom\'e materials both follows a quadratic scaling with similar values regardless of applied pressure thereby suggesting a universal origin of AHE in these materials. \textbf{a} from Ref. \cite{yang2020} and \textbf{b}, \textbf{c} from Ref. \cite{yu2021}.}
	\label{fig:3}
\end{figure*}

While the initial ARPES measurements have established the $\mathbb{Z}_2$ topological metal, ARPES has also been shown to reveal topological surface states as well as flat bands in CsV$_3$Sb$_5$ \cite{hu2021}, with implications on the electron correlation phenomena. Higher resolution ARPES has visualized the temperature-driven band renormalization \cite{liu2021} and the momentum dependence of the CDW gap (largest near M, gapless or small gap near the Dirac cones) in CsV$_3$Sb$_5$ \cite{wang2021, nakayama2021, luo2021} and in KV$_3$Sb$_5$ (in addition to electron-phonon coupling signatures) \cite{luo2022}; whereupon they indicate how the CDW is driven by the scattering of electrons between neighboring M saddle point. Optical conductivity and reflectivity \cite{zhou2021, uykur2021} measurements have also supplemented ARPES measurements by revealing more details on the CDW gap near the M saddle points through transfers of spectral weight. It was shown that the CDW order of the system induces a reconstruction of the Fermi surface pockets through a study of the Fermi surface topography from quantum oscillations \cite{ortiz2021(2)}. Two different types of conventional vHs, p-type (pure) and m-type (mixed), arise at the M point from the intrinsic particle-hole asymmetry of the kagom\'e net with different filling fractions at the Fermi surface, determined by the weights of the distinct sublattice flavors in the system. Among the four vHs near the M point, two  (p-type with filling fraction 5/12) are conventional and found to be very close to the Fermi surface, whereas one of them is a conventional m-type and slightly above the Fermi level. Both the p-type vHs and the Fermi surface are observed in ARPES measurements to be perfectly nested in CsV$_3$Sb$_5$ \cite{kang2022} and in RbV$_3$Sb$_5$ \cite{cho2021}. The remaining vHs, found in ARPES measurements of CsV$_3$Sb$_5$ \cite{hu2021(2)}, is a higher-order p-type which has a flat dispersion along one direction and density-of-states with a power-law divergence which may lead to nematic order. From previous theoretical studies unique to kagom\'e lattices, phenomena such as unconventional superconductivity and the CDW order (possibly with a chiral nature) can be explained through the occurrence of such Fermi surface nesting and vHs conditions \cite{nandkishore2012, kiesel2013}. An experimental observation of quantum well states residing on the top two kagom\'e layers which result from quantum confinement \cite{cai2021} offers a possibility for the discrepancies between the ARPES measurements of the vHs and DFT calculations and an absence of a $k_z$ dispersion.

Seebeck and Nernst measurements have been performed on CsV$_3$Sb$_5$ to investigate how band topology near the Fermi level plays a role in the thermoelectric properties of the system by probing of the entropy flow of carrier quasiparticles which has increased sensitivity to the Berry curvature \cite{gan2021, chen2021(4)}. The magneto-Seebeck signal is concomitant with the CDW transition temperature. Below $\sim$30-35 K, within the CDW phase, both Seebeck and Nernst signals showcase prominent quantum oscillations (Figs. \ref{fig:2}b-c), which are sensitive to the Fermi surface topology. At $\sim$30-35 K, the Seebeck signal reaches peaks and there is an emergence of a large anomalous Nernst effect (ANE) (Fig. \ref{fig:2}d), which is solely sensitive to Berry curvature near the Fermi level, a prominent indicator of nontrivial topology. Due to their common origin of Berry curvature, the ANE is intimately related to the anomalous Hall effect (AHE) which is discussed in Section \ref{sec:anomalous}. They also observe an additional quantum oscillation frequency in their signals at high magnetic field, which they attribute to the magnetic breakdown across two smaller Fermi surfaces originating from a Dirac band with an associated CDW gap of $\sim$20 meV. The transport is dominated by the small Fermi pockets with small effective masses near the M point of the multiband system, where the topological Dirac bands are found, as previously detailed via ARPES measurements. These features highlight the importance of nontrivial band topology on the transport properties of the system. It has also been observed that the longitudinal thermal conductivity deviates from the Wiedemann-Franz law, whereas the thermal Hall conductivity remains consistent, largely due to possible contributions from charge-neutral carriers \cite{zhou2021(3)}.

\section{Anomalous Transport Response} \label{sec:anomalous}
The anomalous Hall effect (AHE), where charge carriers acquire a velocity transverse to an applied electric field, has been previously studied in ferromagnetic-ordered systems or in those with magnetic field to break time-reversal symmetry. The AHE signal can be distinguished from that of the ordinary Hall effect because it is not proportional to the magnitude of the applied magnetic field. Intrinsic AHE refers to the acquisition of a transverse momentum due to the electronic structure of a material endowing a large Berry curvature, and has been found in topological materials such as Co$_3$Sn$_2$S$_2$ \cite{liu2018,wang2018}. When the Dirac points are located near the Fermi level, the breaking of time-reversal symmetry opens an energy gap at these points, which leads to the generation of large Berry curvature and results in phenomena such as the AHE and the ANE (in magneto-thermoelectric measurements). On the other hand, extrinsic AHE refers to the skew scattering, the scattering off of structural defects or magnetic impurities, which can also result in a transverse electron momentum. 

As shown in Fig. \ref{fig:3}, the extracted values of the AHE conductivity ($\sigma_{xy}^{\text{AHE}}$) for AV$_3$Sb$_5$ compounds reach extremely high values of $>10^4$ $\Omega^{-1}$cm$^{-1}$ and large values of anomalous Hall ratio of $\sim$ 1.8\%, both of which are one order of magnitude larger than Fe. The authors of Ref. \cite{yang2020} attribute this giant AHE to the enhanced skew scattering that scales quadratically, rather than linearly, with the longitudinal conductivity ($\sigma_{xx}$) as a result of highly conductive Dirac quasiparticles within the frustrated magnetic sublattice or due to scattering off of tilted spin clusters, local groups of coupled spins which can result from the magnetic atoms in kagom\'e nets. Through this new mechanism, the AV$_3$Sb$_5$ family of materials would open new frontiers for AHE in terms of conceivably reaching an anomalous Hall angle of 90$^\circ$ and for the closely-related spin Hall effect (SHE) in spintronic applications. On the other hand, $\mu$SR and rotation measurements \cite{kenney2021} on KV$_3$Sb$_5$ seem to dispel this possible theoretical explanation for the AHE in this family of kagom\'e metals as there is no evidence for the existence of V local moments that would underlie the existence of these tilted spin clusters. It is therefore likely that the AHE has a complex origin possibly from both alternative skew scattering (lattice geometry intertwined with CDW, for example) and Berry curvature \cite{yu2021}, possibly from the opening of a topological energy gap at the Dirac points from an unconventional CDW order. Efforts are being taken to measure the AHE in nanoflakes \cite{zheng2021}.

Measurements of the ANE signals from Nernst measurements in CsV$_3$Sb$_5$ which decrease at higher temperatures are shown in Fig. \ref{fig:2}d \cite{gan2021, chen2021(4)}. Thermoelectric effects like the ANE are more sensitive to the Berry curvature than electronic effects such as the AHE, and therefore provide stronger relevant information on the band structure. It is proposed from Ref. \cite{gan2021} that ambipolar transport of carriers from compensated bands transpires as evidenced by the Hall coefficient from AHE changing sign at the temperature where the Seebeck signal peaks and the large ANE emerges. Authors in Ref. \cite{chen2021(4)} proposed that the emergence of ANE at this lower temperature could be an indication of the time-reversal-symmetry breaking, explained by the theorized chiral flux phase, as supported by $\mu$SR measurements.

\section{Multitude of coexisting orders} \label{sec:STM}
\begin{figure}[ht!]
	\centering
	\includegraphics[width=\linewidth]{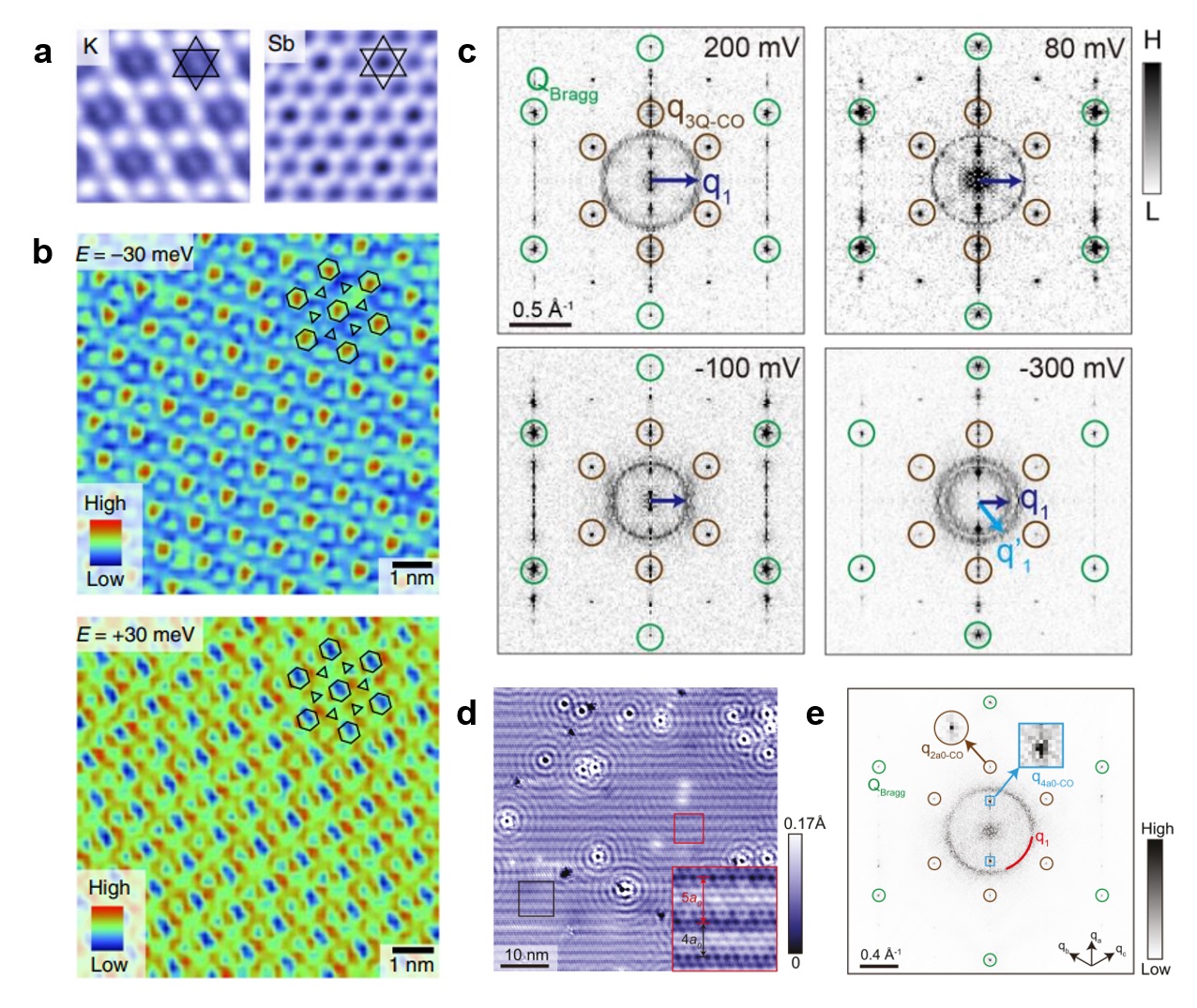}
	\caption{\textbf{STM measurements of charge order.} \textbf{a)} Atomically resolved topographic images of K and Sb surfaces on KV$_3$Sb$_5$ and \textbf{b)} $dI/dV$ imaging for the Sb surface at $\pm$30 meV. \textbf{c)} Fourier transform of $dI/dV$ maps on Sb surface of CsV$_3$Sb$_5$ showing scattering vector $\textbf{q}_1$, atomic Bragg peaks (green circles) and charge order $\textbf{q}_{2a_0-\text{CO}}$ (brown circles). \textbf{d)} STM topograph at 4.5 K taken over an Sb-terminated surface with \textbf{e)} corresponding Fourier transform, indicating Bragg peaks (green circles), $\textbf{q}_{2a_0-\text{CO}}$ peaks (brown circles) and $\textbf{q}_{4a_0-\text{CO}}$ peaks (blue squares). $2a_0$-CO indicates a static charge order with a $2a_0$ period in all directions and similarly for $4a_0$. \textbf{a}, \textbf{b} from Ref. \cite{jiang2021} and \textbf{c}-\textbf{e} from Ref. \cite{zhao2021(2)}.}
	\label{fig:4}
\end{figure}

STM serves as a versatile technique that visualizes the electron distribution on the surface of materials with high spatial resolution and possibilities of an applied magnetic field, thereby allowing the discovery and tunability of many-body phenomena in kagom\'e lattice systems. The first high-resolution STM measurements of KV$_3$Sb$_5$ \cite{jiang2021} (Fig. \ref{fig:4}a for topographic images of K and Sb surfaces) reveal a robust in-plane $2\times2$ superlattice structure and unconventional charge ordering through intensity reversals (Fig. \ref{fig:4}b) of the charge modulation pattern upon the opening of the energy gap near the Fermi level. This unusual charge ordering was found to be universal among the three materials of the AV$_3$Sb$_5$ family, where STM measurements of RbV$_3$Sb$_5$ \cite{shumiya2021} and CsV$_3$Sb$_5$ \cite{wang2021(2)} show similar results. The wavevector of the charge modulation matches the nesting condition between the van Hove singularities that are found at the M points of the Brillouin zone. The authors explain the unconventional CDW order as a chiral CDW in a kagom\'e lattice, whereupon the chirality bestows an intensity anisotropy along different directions (clockwise/counter-clockwise) in the charge modulation vector peaks. This handedness can be tuned or switched when a magnetic field is applied, indicating a time-reversal symmetry breaking state. Chiral CDWs would have major implications for unconventional (nodal) superconductivity, correlation-driven AHE and chiral Majorana zero modes.

Subsequent STM measurements have established a series of correlated electron states in CsV$_3$Sb$_5$ which persist upon cooling below the superconducting transition temperature \cite{zhao2021(2)}. In particular, as shown in Fig. \ref{fig:4}c, aside from atomic Bragg peaks, the $2\times2$ superstructure seen from a Fourier transform of the $dI/dV$ maps give rise to a static charge order with a $2a_0$ period along all three lattice directions (called $2a_0$-CO). At lower temperatures ($\sim$50 K) (Figs. \ref{fig:4}d-e), another periodic modulation with a $4a_0$ wavelength (called $4a_0$-CO) propagates unidirectionally, breaks rotation symmetry and possesses strongly anisotropic scattering -- reminiscent of a charge stripe modulation. This temperature coincides with that of the onset of resistivity anisotropy in electric transport measurement. The $4a_0$ stripe modulation is attributed to a strong energy-dependent orbital renormalization of the V kagom\'e bands. There are other supporting evidence of this nematic $1\times4$ superlattice modulation \cite{shumiya2021, wang2021(2), wang2021(6), zhao2021(2)} in RbV$_3$Sb$_5$ and CsV$_3$Sb$_5$. Further measurements \cite{li2021(2)} reveal that there is an intensity anisotropy of one $2a_0$ CDW direction (of the three) in comparison to the other two which would point to an intrinsic rotation symmetry breaking where the symmetry is broken from $C_6$ to $C_2$. The authors of Ref. \cite{li2021(2)} also report an absence of magnetic field dependence of the CDW phase, presenting a conflicting observation to Ref. \cite{jiang2021} and questioning the robustness of these states in STM. In contrast, studies in Ref. \cite{li2021(4)} involving high-resolution x-ray diffraction, STM and scanning transmission electron microscopy (STEM) suggest that the $4a_0$ modulation emerges only at the surface and is not a bulk phenomena. It is explained that the breaking of bulk rotation symmetry from $C_6$ to $C_2$ from the three-dimensional $2\times2\times2$ CDW order drives the nematic $1\times4$ orders observed in the bulk and at the surface. STM measurements with spin-polarized tips observe the 2$\times$2 CDW and $4a_0$ stripe orders, but there is an absence of a spin texture, up to experimental resolution, which would correspond to the theoretically-proposed chiral flux phase, a chiral charge order or a spin density wave order in CsV$_3$Sb$_5$ \cite{li2022}. Another intriguing feature is the presence of a $4a_0/3$ bidirectional roton pair density wave in CsV$_3$Sb$_5$ \cite{chen2021(2)}, the latter of which possesses similarities to the phenomenology of cuprate high-temperature superconductors. Measurements of the superconducting gap suggest multiband superconductivity with an s-wave superconductivity order parameter \cite{xu2021}.

\section{Theoretical works and calculations} \label{sec:theory}
Several theoretical proposals with associated calculations have been put forward to explain the exotic CDW ordering observed in the family of kagom\'e metals AV$_3$Sb$_5$ (A = K, Cs, Rb). Extensive DFT calculations have also shown the electronic correlations and magnetic properties in the normal state \cite{zhao2021(3)}. Below the CDW transition temperature, studies involving first-principles calculations \cite{tan2021} and machine-assisted statistical learning approaches \cite{mertz2021} suggest that the $2\times2$ CDW ground state of the kagom\'e lattice is an inverse Star of David structure with the transition being driven by a Peierls instability. This supports the conclusion of the hard x-ray scattering measurement in Ref. \cite{li2021}, that the electron-phonon coupling in the system is too weak to rationalize conventional superconductivity and there is an absence of an acoustic phonon anomaly. The three-dimensional inverse Star of David structure is shown numerically \cite{miao2021} to be consistent with the x-ray scattering measurement of a $2\times2\times2$ superstructure and with scanning tunneling spectroscopy measurements displaying a breaking of the sixfold symmetry. A $\pi$-phase shift between adjacent kagom\'e layers breaks the $C_6$ symmetry which modifies the CDW peak intensities through a weak first order phase transition via a coupling between lattice distortions and the CDW. One study \cite{wang2021(7)} performs a DFT calculation to show that the CDW instability and associated structural transition are accompanied by the stabilization of quasimolecular states, which leads to the opening of CDW gaps and suggests that the Jahn-Teller effect drives the CDW order.

Studies have also been done to determine the strength of the electron-phonon coupling in these materials and whether phonons play a large role in the superconductivity. A computational study of the system under pressure would indicate conventional superconductivity, with a partial suppression at ambient and low pressures due to the magnetism from vanadium atoms \cite{zhang2021(2)}. Conversely, first-principles calculations of phonon instabilities due to possible structural phase transition near the CDW transition temperature have been performed \cite{subedi2022} and reveal phonon softening at the M and L points which can lead to an emergence of an inverse Star of David pattern \cite{ptok2021}. Another study \cite{ye2021} reveals that the CDW in the system is mainly driven by a phonon instability and electron-phonon coupling as well as a metastable $1\times4$ phase with dimerization of the V atoms in a twofold symmetric bowtie-shaped pattern. These theoretical studies will help elucidate more on the electron-phonon coupling measurements performed in x-ray scattering, neutron scattering and ARPES measurements.

Another argument for the $2\times2$ charge orders and the emergence of anomalous transport comes from describing the ground state as a chiral flux phase \cite{feng2021, feng2021(2)} where a tight-binding model is used to rationalize an imaginary order parameter for the charge bond order. As mentioned previously, there are numerous bits of evidence that either support or dispel the existence of a chiral flux phase in these materials. A similar model to the chiral flux phase containing vHs with a random phase approximation analysis for the superconducting instabilities may elucidate further information on the role of doping and pressure on the superconductivity of these materials as observed in experiments through Fermi surface reconstruction or vHs tuning \cite{wu2021}. The effect tuning of the vHs near the Fermi level with applied pressure and hole doping on the electronic structure was numerically studied \cite{labollita2021} -- displaying a reconstruction of the Fermi surface with respect to the Sb bands. This was also studied in Ref. \cite{consiglio2021} for applied pressure, showcasing a shifting the p-type and m-type vHs with respect to the Fermi level which may serve as an experimental knob to further explore the suppression of charge orders and the double-dome superconductivity. This is related to the anisotropic compression proposed in Ref. \cite{tsirlin2021}. The saddle points move away from the Fermi level with pressure, and vice-versa for increased hole doping. Additionally, some have suggested the existence of smectic bond-density wave \cite{tazai2021} or time reversal symmetry-breaking density wave \cite{setty2021} mechanisms. An orbital-based mechanism in Ref. \cite{zhou2021(2)} is conceived to connect the small (Chern) Fermi pockets from quantum oscillations measurements and the observation of pair density waves in STM through a one-orbital model on a kagom\'e lattice lightly doped away from the van Hove filling.

Lastly, Landau theory is commonly used to explain the formation of complex orders due to the presence of vHs on the hexagonal lattice and electronic instabilities leading to different types of density waves. Development of these models may serve to predict new phenomena that either showcase traces of occurrence in realized experiments or await future experimental verification in these materials. Refs. \cite{lin2021, lin2021(2), lin2021(3)} demonstrate a rich Haldane-model phase diagram of the multi-component hexagonal (3Q) charge density waves, where they find trivial and Chern insulator phases with real or imaginary orders (like the chiral flux phase). In particular, the coupled mean field theory of spin and charge orders may describe the topological charge density wave in STM measurements as being identified with the complex orders of the Chern insulator phase and the rotation symmetry breaking results from secondary orders of the complex order ground states. Subsequent work along these lines showcases higher-order topological insulator states and superconducting pairing resulting from Pomeranchuk fluctuations which may explain the double-dome superconductivity and large separation in energy scales between superconductivity and CDW. Ref. \cite{park2021} builds upon this approach, using parquet renormalization theory, to reveal the enhanced electronic instabilities by vHs near the M saddle points of the Brillouin zone inducing scattering amongst these points, as observed through ARPES measurements. Similar ideas have emerged for graphene and more recently, magic-angle bilayer graphene. Their analysis provides constraints on the interpretation of several experiments with a focus on the density wave order. Ref. \cite{christensen2021} classify the different types of multi-Q CDW orders to study the leading instabilities of these kagom\'e systems. Ref. \cite{denner2021} develops a theory for the CDW formation using Ginzburg-Landau formalism, finding it is energetically preferred that a nematic chiral charge order is the ground state of the system.

\section{Outlook} \label{sec:outlook} 
From these rapid developments of experimental and theoretical results, the AV$_3$Sb$_5$ (A = K, Rb, Cs) class of kagom\'e metals consists of an all-in-one system allowing an exciting opportunity to study the interplay between nontrivial topological band physics, geometric frustration, and unconventional superconductivity. While kagom\'e materials have been extensively studied for decades as they are natural platforms for frustration-driven exotic quantum phases, dispersionless flat bands and linearly-dispersive Dirac energy bands, the AV$_3$Sb$_5$ set has seen a surge of interest due to the presence of superconductivity, presumably of an unconventional nature, in conjunction with the presence of competing instabilities. Accordingly, these materials have been examined using the vast majority of experimental probes: angle-resolved photoemission spectroscopy, bulk electrical and thermal transport, x-ray and neutron scattering, high pressure studies, scanning tunneling microscopy, pump-probe measurements, muon spin spectroscopy, NMR, optical conductivity and reflectometry, ultrafast coherent phonon spectroscopy, Raman scattering, point-contact spectroscopy and others. In spite of abundant experimental research, there are puzzling questions that remain unanswered. The different temperature scales among these different characterization studies ($\sim$30-35K for thermal transport, $\sim$70 K for muon spin spectroscopy ($\mu$SR), $\sim$50 K for Raman scattering and $\sim$60 K for anisotropic superconducting critical fields), which all occur below the CDW transition temperature of $\sim$94 K in CsV$_3$Sb$_5$, require further theoretical and experimental investigations in order to mediate between them.

STM measurements have shown that rotational symmetry breaking is an intrinsic property in one of the kagom\'e metals, CsV$_3$Sb$_5$, via newly-observed symmetry-broken electronic states. The latter includes an unidirectional $4a_0$ charge order which will trigger more theoretical efforts to explain its origin within kagom\'e lattices, but also in other types of materials. Future questions present themselves: how do the relations between these different phases evolve with temperature? How are they modified with different dopings that tune the state away from the van Hove saddle point? Future STM measurements involving temperature, energy and doping dependencies would address questions into the coherence of the CDW, on the tunability of the CDW state, on the superconducting gap functions and on the nature of the scattering involved within this material. The ability to tune away from the van Hove saddle point as well as the multi-orbital nature of AV$_3$Sb$_5$ provide an exciting opportunity to explore in more detail.

There is an evident resemblance between the electronic phenomena that occur in this system to that of unconventional high-temperature superconductors such as the cuprates. As the AV$_3$Sb$_5$ is a cleaner system with no doping necessary, studies into this material will help shed light on the enigmatic mechanisms that occur in cuprate systems, such as the relation between the CDW phase and the superconductivity. For example, AV$_3$Sb$_5$ also offers an arena to study multiple charge orders that compete with superconductivity as well as nematicity and band-dependent gapping of different regions of the Fermi surface. Studies at high pressures reveal temperature-pressure phase diagrams with a double-dome feature and hints of quantum criticality at the second superconducting dome peak ($P_{C2}$) that resemble those of several unconventional superconductors. Future studies into AV$_3$Sb$_5$ should address the CDW nature in the region between the two superconducting domes and the question of quantum criticality near the first superconducting dome ($P_{C1}$) at ultra-low temperature. 

Lastly, while most studies have shown that the superconductivity may be unconventional through thermal transport, the presence of double superconducting domes, point-contact spectroscopy measurements and pair density waves in STM, the true pairing mechanism and symmetry is still up for debate. Other measurements such as that of the magnetic penetration depth and of multiband superconductivity with a sign-preserving order parameter nuance the idea of unconventional superconductivity and instead favor s-wave superconductivity. More experimental evidence is required to conclusively determine the nature of the superconductivity of the compound. The nontrivial topological band structure may engender the formation of zero energy modes related to Majorana bound states modes within the vortex cores of inherent proximitized superconducting surface states. Evidence for the latter has been observed through topological CDWs, possibly with chiral flux, which give rise to other instabilities such as the pair density wave (PDW) and in signatures of spin-triplet pairing and edge supercurrents. Notwithstanding the unique combination of unconventional superconductivity and nontrivial topology in a single material, the presence of a large anomalous Hall effect (AHE), proposed as a product of the chiral CDW in the system, also pushes the quantum material into the application-based realm of quantum information science as Majorana zero modes and of energy-relevant technologies such as spintronics. Future studies should tackle the ongoing inquiries on the unconventional superconductivity and on the origin of the large AHE in order to drive the developments of applications with AV$_3$Sb$_5$ materials. As the material is an easily exfoliable kagom\'e metal, the potential for future work beyond what was presented in this brief review is high in the form of studies in thin films and heterostructures devices similar to the transition metal dichalcogenides. 

The recent surge of research into AV$_3$Sb$_5$ kagom\'e metals on both experimental and theoretical fronts has led to an abundance of new viewpoints into exciting topics such as strong electron correlation, unconventional superconductivity and nontrivial band topology. Studies into these materials are currently ongoing with improved experimental techniques, different experimental probes and novel theoretical models to account for the phenomena observed in these kagom\'e metals. The desire to find materials that possess unique topological phases and incorporate geometric frustration and strong electron correlation has been revitalized with the discovery of these vanadium-based kagom\'e compounds, particularly noteworthy is the discovery and characterization of homologous families of compounds \cite{shi2021, yang2021, yin2021(3), yang2021(2)} that show an absence of superconductivity and CDW order, but possible Dirac nodal lines near the Fermi level. Altogether, further research into AV$_3$Sb$_5$ has a solid foundation and promising prospects ahead for meaningful scientific insights that may resolve some of the outstanding issues in different fields of condensed matter physics.

\begin{acknowledgments}
T.N. and M.L. acknowledge the support from U.S. Department of Energy (DOE), Office of Science (SC), Basic Energy Sciences (BES), award No. DE-SC0020148. M.L. acknowledges the support from NSF DMR-2118448, the Norman C. Rasmussen Career Development Chair, and support from Dr. R.A. Wachnik. 
\end{acknowledgments}

\bibliographystyle{apsrev4-1}
\bibliography{references.bib}


\end{document}